# Evidence of Low-Energy Singlet Excited States in the Spin-1/2 Polyhedral Clusters {Mo$_{72}$V$_{30}$} and {W$_{72}$V$_{30}$} with Strongly Frustrated Kagomé Networks

T. Kihara[1*], H. Nojiri[1], Y. Narumi[2], Y. Oshima[3], K. Kindo[4], C. Heesing[5], J. Schnack[5], and A. Müller[6]

[1]Institute for Materials Research, Tohoku University, Katahira 2-1-1, Sendai 980-8577, Japan
[2]Center of Advanced High Magnetic Field Science, Osaka University, Toyonaka 1-1, Osaka 560-0043, Japan
[3]Condensed Molecular Materials Laboratory, RIKEN Cluster for Pioneering Research, Hirosawa 2-1, Wako, Saitama 351-0198, Japan
The Institute for Solid State Physics, The University of Tokyo, Kashiwanoha 5-1-5, Kashiwa 277-8581, Japan
[5]Fakultät für Physik, Universität Bielefeld, Postfach 100131, D-33501 Bielefeld, Germany
[6]Fakultät für Chemie, Universität Bielefeld, Postfach 100131, D-33501 Bielefeld, Germany

*E-mail address: t_kihara@imr.tohoku.ac.jp

## Abstract

Magnetization, specific heat, and electron spin resonance (ESR) measurements are carried out to clarify the low-energy excitations for the $S$ = 1/2 polyhedral clusters {Mo$_{72}$V$_{30}$} and {W$_{72}$V$_{30}$}. The clusters provide unique model systems of Kagomé network on a quasi-sphere. The linear field variation of magnetization at low temperatures indicates that the ground state is singlet for both clusters. The temperature and the magnetic field dependence of specific heat shows a distinct difference between two clusters with differing structural symmetries. In {W$_{72}$V$_{30}$} with $I_h$ symmetry, the existence of the several tens of low-energy singlet excited states below the lowest triplet excited state is revealed. The specific heat of the slightly distorted {Mo$_{72}$V$_{30}$} with $D_{5h}$ symmetry leads to a drastic decrease of the singlet contribution, which is consistent with the partial lifting of the frustration and the decrease of degenerated low-energy states. The singlet excitation existence is confirmed further by the temperature dependence of the ESR spectra. Comprehensive experimental studies have demonstrated unique low-energy excitations of the spherical Kagomé networks and their sensitivity to the cluster symmetry.

# I. Introduction

In geometrically frustrated spin systems, conventional long-range orderings are prevented because of the enormous degeneracy of the ground state. In such cases, nontrivial ground states such as spin-liquid states or unusual ordered states with exotic order parameters are expected.[1] The Kagomé lattice, a unique network of corner-sharing triangles, has attracted much attention in recent decades as a prominent ingredient for exploring novel magnetic ground states in highly frustrated spin systems. The classical antiferromagnetic Heisenberg model on the Kagomé lattice holds a macroscopic degeneracy of ground states even at $T = 0$, because of its corner-sharing geometry. The resultant zero-energy excitations prevent the conventional long-range ordering. Several theoretical works about this model predict that the so-called $\sqrt{3} \times \sqrt{3}$ order, in which the three classical spins of a triangle are fixed at relative angles of 120° to each other in a plane, is realized as the ground state at low temperatures.[1–3]

When moving from the classical to the quantum Kagomé Heisenberg antiferromagnets with $S = 1/2$, a completely different low-energy state can be realized. Because of the existence of strong quantum fluctuations, extensive singlet excitations are created below the first excited triplet state.[1,4] It has been proposed theoretically that such highly degenerated and highly fluctuating states lead to the realization of unusual ground states such as the resonating valence bond (RVB) state[5,6] and the spin-liquid state.[7–9] To date, however, only a few experimental realizations have been confirmed because of the difficulty in obtaining a proper model system with a Kagomé lattice network with high symmetry even at low temperatures.[10]

Herein, we present a molecular Kagomé network: a spherical cluster of $S = 1/2$ spins that can be regarded as zero-dimensional analogue of the Kagomé lattice. This study examines two systems $\{Mo_{72}V_{30}\}$[11,12] and $\{W_{72}V_{30}\}$,[13,14] where the $S = 1/2$ spins of $V^{4+}$ form an icosidodecahedron. This network consists of 20 corner-sharing triangles and 12 pentagons, which makes it an ideal model system for spherical Kagomé networks. A series of spherical Kagomé clusters such as $\{Mo_{72}Fe_{30}\}$ ($S = 5/2$)[15] and $\{Mo_{72}Cr_{30}\}$ ($S = 3/2$) also exists,[16] with evidence of strong frustration found in the magnetic properties.[17–20] A modified cluster such as {Mo$_{75}$V$_{20}$} has also been examined, which is regarded as a saw-tooth chain with periodic boundary conditions.[21]

For {Mo$_{72}$V$_{30}$}, the temperature dependence of the magnetic susceptibility indicates a singlet ground state.[11,12] Müller *et al.* simulated the magnetic susceptibility at temperatures higher than 120 K using the quantum Monte Carlo method for the isotropic antiferromagnetic Heisenberg model with $I_h$ symmetry, where the spin of each V$^{4+}$ ion is coupled to its four nearest-neighbors via intramolecular antiferromagnetic exchange interactions, with estimated the nearest-neighbor exchange interaction $J$ = 245 K.[11] Nevertheless, this model cannot reproduce the experimentally obtained results at lower temperatures. Then, Kunisada *et al.* introduced a slightly distorted icosidodecahedron to resolve this discrepancy between the experimental and the theoretical results. The proposed antiferromagnetic Heisenberg model with $D_{5h}$ symmetry successfully reproduces the experimentally obtained magnetic susceptibility of {Mo$_{72}$V$_{30}$} for all temperature ranges.[22,23] The energy gap separating the ground state and the first triplet excited state is calculated as 12.3 K (9.4 T) for {Mo$_{72}$V$_{30}$}.

In {W$_{72}$V$_{30}$}, the temperature dependence of the magnetic susceptibility is well-reproduced by the isotropic antiferromagnetic Heisenberg model with $I_h$ symmetry for the exchange energy of $J$ = 115 K.[23] The low-temperature magnetization curve is also calculated using this model. It shows several steps for the level crossing. However, the experimentally obtained *M-H* curve measured at 0.5 K in the pulsed magnetic field up to 50 T exhibits no such staircase behavior,[24] although the first step is expected to appear at 19.2 T in the calculation.[23] Schnack *et al.* theoretically assessed the distribution of nearest-neighbor exchange couplings to explain the discrepancy between theory and experiment for *M-H* curves.[24] However, the origin of the experimentally obtained monotonic *M-H* curve for {W$_{72}$V$_{30}$} with no staircase behavior is not clear at present.

As described in this paper, we conducted a comprehensive study by combining magnetization, specific heat, and electron spin resonance (ESR) measurements using both the steady and the pulsed high magnetic fields on {Mo$_{72}$V$_{30}$} and {W$_{72}$V$_{30}$}, to investigate the characteristics of the low-energy excitations.

Systematic study uncovered the unique singlet excitations and their dependences on symmetries in a Kagomé system.

This paper is organized as follows. Section II briefly presents the experimental details. Then, the experimentally obtained results of magnetization, specific heat, and ESR measurements are provided in Sec. III. Section IV presents a summary of our findings and identifies the remaining open questions.

## II. Experimental methods

Polycrystalline samples of $\{Mo_{72}V_{30}\}$ and $\{W_{72}V_{30}\}$ were synthesized using the procedures explained in an earlier report[14]. Magnetization measurements in the high magnetic fields of up to 27 T for $\{Mo_{72}V_{30}\}$ and up to 50 T for $\{W_{72}V_{30}\}$ were taken using nondestructive pulse magnets installed at the Institute for Materials Research (IMR) of Tohoku University and at the Institute for Solid State Physics (ISSP) of the University of Tokyo.[25,26] The magnetization data were collected using the induction method with coaxial-type pick-up coils. The samples were immersed in the liquid helium to suppress the temperature variation during the measurements. Specific heat measurements in the steady field up to 15 T were conducted using thermal relaxation technique with a homemade high sensitivity probe at the High Field Laboratory for Superconducting Materials (HFLSM) of Tohoku University.[27] The ESR measurements in the pulsed fields were taken using a terahertz ESR apparatus (TESRA) installed at the IMR at the frequencies of 190 and 405 GHz and at the temperatures of 1.5-70 K.[28]

## III. Results and Discussion

### A. Magnetization

We first took the magnetization measurements for $\{Mo_{72}V_{30}\}$ and $\{W_{72}V_{30}\}$ using the pulsed magnetic fields, to examine the ground state characteristics. Figures 1(a) and (b) show the *M-H* curves of $\{Mo_{72}V_{30}\}$ and $\{W_{72}V_{30}\}$ measured respectively at 0.5 K and 1.4 K. For both compounds, the magnetization increases linearly accompanied with the paramagnetic contribution below 5 T. The experimental results plotted as solid curves in Figs. 1(a) and (b) are well-fitted by a sum of the Brillouin function for $S = 1/2$ and the linear term shown as dashed curves in the figures. It is known that some extra amount of vanadyl ions inevitably remains through the synthetic process of these types of polyoxometalate clusters.[29,30] For that reason, the paramagnetic contribution can be attributed to the presence of free spins from the $V^{4+}$ impurities. From the comparison of the paramagnetic contribution with the linear term in the fitted *M-H* curves, it is estimated that about 0.88 and 1.07 impurities exist, respectively, per {Mo$_{72}$V$_{30}$} cluster and {W$_{72}$V$_{30}$} cluster. Hysteresis loops are apparent below 5 T in the *M-H* curves in Figs. 1(a) and (b). In pulsed field experiments of magnetic molecules, when thermal relaxation rate of spin is comparable with the magnetic field sweep rate of the pulsed field, it gives rise to an extrinsic hysteresis loop in a *M-H* curve.[31] Consequently, these hysteresis loops below 5 T in the Figs. 1(a) and (b) are not attributable to magnetic ordering.

Removal of the paramagnetic impurity contribution reveals intrinsic magnetization for both compounds, which are observed up to 27 T for {Mo$_{72}$V$_{30}$} and 49 T for {W$_{72}$V$_{30}$} and which are shown as the gray solid curves in Figs. 1(a) and (b). The linear magnetizations crossing the origin indicates that the ground state is singlet in both compounds, which is consistent with the results of the magnetic susceptibility.[12–14] The coefficients of these linear magnetizations are estimated as 0.0493 $\mu_B$/T (0.0275 emu/G/mol) for {Mo$_{72}$V$_{30}$} and 0.0639 $\mu_B$/T (0.0357 emu/G/mol) for {W$_{72}$V$_{30}$}, respectively. If we assume linear field dependence of magnetization even for higher fields, the saturation fields are estimated respectively as 609 T for {Mo$_{72}$V$_{30}$} and 469 T for {W$_{72}$V$_{30}$}. Although these are rough estimations, we can claim that the interaction between spins in {Mo$_{72}$V$_{30}$} is stronger than that in {W$_{72}$V$_{30}$}, which is consistent with the nearest-neighbor exchange interactions ( $J$ = 245 K for {Mo$_{72}$V$_{30}$} and $J$ = 115 K for {W$_{72}$V$_{30}$}) estimated by comparison the model calculation with the experimental results of the magnetic susceptibility.[23]

Next, we discuss the ground state. Magnetization measurements at low temperatures are useful to investigate the field variation of the ground states. When the total spin quantum number of the ground state changes from $S$ to $S + 1$ at a level crossing, a magnetization step from $M = 2S$ to $M = 2(S + 1)$ is observed. In the theoretically calculated $M$-$H$ curves for a cluster with high symmetry, the steps attributable to the level crossing are almost uniformly spaced.[23,24,32,33] When the cluster is slightly distorted, low-lying states are partially lifted, and therefore, the level crossing fields are changed. The first steps attributable to level crossing between the singlet ground state and the triplet excited state are estimated at 9.4 T for {Mo$_{72}$V$_{30}$} and at 19.2 T for {W$_{72}$V$_{30}$}[23] [see the blue dashed lines in Figs. 1(a) and (b)]. Figures 1(a) and (b) show that no staircase like behavior exists in either compound. Here, the measurement temperature is sufficiently low. The effect of thermal excitation is excluded as the origin of the monotonic $M$-$H$ curves. Therefore, we must consider the hybridization of states to explain the disappearance of the steps in the $M$-$H$ curves.

Such monotonic $M$-$H$ curves were observed not only in the $S = 1/2$ system, but also in other spherical Kagomé clusters with large spin such as {Mo$_{72}$Fe$_{30}$} ($S = 5/2$) and {Mo$_{72}$Cr$_{30}$} ($S = 3/2$).[34] To date, several works have proposed distribution of the exchange interactions, Dzyaloshinsky-Moriya (DM) interactions, or intermolecular interactions to realize the hybridization of states.[23,24,34–36] For the $S = 5/2$ and $S = 3/2$ systems, the $M$-$H$ curves at low temperatures are well-reproduced by the classical Heisenberg model with the randomly distributed nearest-neighbor exchange interactions.[34] In cases of $S = 1/2$ systems such as {Mo$_{72}$V$_{30}$} and {W$_{72}$V$_{30}$}, the small magnetization steps remain in the theoretical calculations.[24] Recently, Heesing[37] and Fukumoto et al.[35] independently introduced the DM interactions to explain the disappearance of the steps, and obtained the linear like magnetization. Figure 1(c) presents an example in which an almost linear dependence of $M$ vs. $H$ can result from strong DM interactions.[37] Consequently, the DM interaction represents a possible origin of the monotonic $M$-$H$ curves of {Mo$_{72}$V$_{30}$} and {W$_{72}$V$_{30}$}. Because the spins in a cluster are coupled with each other via the -O-M-O- (M = Mo or W) bridge, the magnitude of the DM interaction depends on the spin-orbit interaction of the M element. Although the spin-orbit coupling constant of the W ion is several times larger than that of the Mo ion,[38] the similar $M$-$H$ curves are obtained for both compounds. For {Mo$_{72}$V$_{30}$}, the structural distortion of a cluster also contributes to the distribution of states, as described in the next subsection.

It is also noteworthy that, although the energy gap separating the singlet ground state and the first triplet excited state of {Mo$_{72}$V$_{30}$} is smaller than half that of {W$_{72}$V$_{30}$},[23] the slope of the experimentally obtained *M-H* curve of {Mo$_{72}$V$_{30}$} is smaller than that of {W$_{72}$V$_{30}$}. This relatively small slope of magnetization for {Mo$_{72}$V$_{30}$} cannot be explained by the DM interaction. We here consider the contribution of the S ≥ 1 multiplet ground states to the magnetization. For {W$_{72}$V$_{30}$} with $I_h$ symmetry, the three steps attributable to level crossings among $S$ = 0, 1, 2, and 3 states are theoretically estimated below 50 T (i.e. the magnetization reaches 6 μ$_B$ at 50 T).[23] On the other hand, for {Mo$_{72}$V$_{30}$} with $D_{5h}$ symmetry, the two steps attributable to level crossings among $S$ = 0, 1, and 2 states are theoretically estimated below 50 T (i.e. the magnetization reaches 4 μ$_B$ at 50 T). Because not only the first level crossing but also the second and the third level crossings contribute to the magnetization below 50 T, the relatively small slope of magnetization for {Mo$_{72}$V$_{30}$} indicates that the low-lying states are partially lifted by the structural distortion of a cluster, which is consistent with the results of the specific heats, as described in the next subsection.

Herein, we present a short summary of findings obtained from magnetization measurements. The ground state is singlet for both {Mo$_{72}$V$_{30}$} and {W$_{72}$V$_{30}$}. The strong DM interaction is a possible origin of the experimentally obtained monotonic *M-H* curves of {Mo$_{72}$V$_{30}$} and {W$_{72}$V$_{30}$}.

### B. Specific heat

Next, we performed specific heat measurements at low temperatures and in magnetic fields to examine the excited states. Figure 2(a) shows the temperature dependence of the specific heats measured at 0, 5, 10, and 15 T for the polycrystalline sample of {Mo$_{72}$V$_{30}$}. Here, the specific heat is presented per unit of molecular weight of M$_{72}$V$_{30}$ (M = Mo and W). Furthermore, the lattice contribution of the specific heat as ascertained from the experimental data of {Mo$_{72}$Fe$_{30}$} at 0 T, was subtracted from the results of {Mo$_{72}$V$_{30}$} to estimate the magnetic contribution. It is noteworthy that the magnetic part of the specific heat of {Mo$_{72}$Fe$_{30}$} is well-reproduced by a conventional two-level Schottky model, and is relatively small above 10 K.[39] Therefore, we estimate the lattice part of the specific heat by fitting the Debye model to the data of the specific heat above 10 K for {Mo$_{72}$Fe$_{30}$}. To remove the extrinsic contribution from the paramagnetic spins in the extra vanadyl ions trapped among molecules as impurities, the temperature and field dependences of Schottky-type specific heats, which can be described as $C_{\mathrm{imp}} = -Ng\mu_\mathrm{B}S\mu_0 H[\partial B_\mathrm{S}(x)/\partial T]$ ($x = g\mu_\mathrm{B}S\mu_0 H/k_\mathrm{B}T$), were also subtracted. Here, the $B_\mathrm{S}(x)$ is the Brillouin function. Also, $N \cong 6.02 \times 10^{23}$ mol$^{-1}$, $\mu_\mathrm{B} \cong 9.274 \times 10^{-24}$ J/T, $\mu_0 = 4\pi \times 10^{-7}$ H/m, and $k_\mathrm{B} \cong 1.38 \times 10^{-23}$ J/K respectively represent Avogadro's number, Bohr magneton, space permeability, and Boltzmann constant. We set the parameters as $g = 2$ and $S = 1/2$. The concentration of impurities of 0.88 per molecule for the {Mo$_{72}$V$_{30}$} sample is estimated from the paramagnetic component in the $M$-$H$ curve, as described in the previous section.

In Fig. 2(a), the temperature dependence of the magnetic specific heat at 0 T shows a peak at around 2.4 K (solid circles). This peak sharpens and increases by application of the magnetic field of 5 T. Subsequently, it broadens and moves slightly toward higher temperatures with increasing magnetic fields up to 15 T. The singlet states are insensitive to the applied magnetic field. Therefore, this field-dependent peak is attributed to contributions from the triplet excitations. Here, we consider a simple model with a singlet ground state and a triplet excited state, with the energy diagram shown schematically in the inset of Fig. 2(a), to provide a qualitative interpretation of the field variation of the peak in the specific heat. In this model, threefold degeneracy of the triplet excited state is split by application of a magnetic field. A level crossing occurs at $B^*$. The magnetic specific heat attributable to the triplet excitations shows a Schottky-type anomaly, where the peak position is determined by the energy gap separating the singlet and the triplet states. When the applied field is lower than $B^*$, the energy gap decreases concomitantly with the increasing field. In this case, the peak of specific heat sharpens and moves toward lower temperatures. When the applied field is higher than $B^*$, the energy gap increases concomitantly with the increasing field. Therefore, the peak of specific heat broadens and moves toward higher temperature with the increasing field. Although this is a rough approximation, the field variation of the experimentally obtained magnetic specific heat can be understood qualitatively considering the existence of low-lying triplet excited states. In the real system, multiple triplet states contribute to the temperature and the field dependence of the specific heat, which can be responsible for the field insensitive peak position and enhancement of the peak height below 5 T. Figure 2(b) portrays the temperature dependences of the entropy calculated from specific heats above 1.6 K. The maximum value of the entropy is only about 3.56 ($\sim Nk_\text{B}$ln1.53) J/mol K at 5 T. This small value of entropy indicates that only a few states contribute to the specific heat at low temperatures. This experimental result implies that the numerous low-lying singlet and triplet states are partly removed by a crystallographic distortion in the icosidodecahedra of {$Mo_{72}V_{30}$}.

The characteristic specific heat behavior is also seen in the theoretical calculations for a cluster with the $D_{5h}$ symmetry by Kunisada et al.,[23] shown as solid and dotted curves in Fig. 2(a). In their calculations, the peak at around 3 K at 0 T moves toward lower temperatures in the applied magnetic fields. It is noteworthy that the position and height of the peak in the calculation are comparable to the experimentally obtained results. However, the field dependence of the peaks mutually differs, which implies that the energy distribution of the triplet excited states is more complicated in the real systems than in the calculations.

Next, we present specific heat results of {$W_{72}V_{30}$} with high symmetry of $I_h$. Figure 3(a) shows the temperature dependence of the magnetic specific heat at 0, 5, 10, and 15 T for the polycrystalline sample of {$W_{72}V_{30}$}. As is the case for {$Mo_{72}V_{30}$}, the lattice contribution and the paramagnetic contribution from the impurity spins were subtracted from the raw specific heat data of {$W_{72}V_{30}$}. Here, the concentration of impurities is 1.07 per {$W_{72}V_{30}$} molecule, as discussed in the preceding section. The obtained magnetic specific heat in Fig. 3(a) differs greatly from that of {$Mo_{72}V_{30}$}. The magnetic specific heat at 0 T monotonically increases concomitantly with increasing temperature up to 10 K. Surprisingly, the magnetic field dependence is nearly absent. This tiny field dependence denies that the specific heat originates from the excited triplets. Another important point is that the absolute values of the specific heat are significantly larger than that of {$Mo_{72}V_{30}$}, which indicates that a large number of excited states exists near the ground state. The result illustrates clearly that there are dense low-energy excited states which are non-magnetic. Figure 3(b) shows the temperature dependence of the magnetic entropy for $\{W_{72}V_{30}\}$ above 0.7 K. The entropy at 10 K is estimated as $S = 35.4 \pm 1.65$ ($\sim Nk_B \ln 70.7^{+15.6}_{-12.7}$) J/mol K at 5, 10, and 15 T. Therefore, it can be inferred that about 71 singlet excited states contribute to specific heat at low temperatures. As described in the *Introduction*, in an ideal Kagomé antiferromagnet with $S = 1/2$, extensive singlet excited states exist below the first triplet excited state because of the strong quantum fluctuations. The large entropy at temperatures lower than 10 K reflects that the icosidodecahedron in $\{W_{72}V_{30}\}$ maintains high symmetry, even at low temperatures, contrary to $\{Mo_{72}V_{30}\}$.

Such high density of singlet excitations at the low energy is apparent in theoretical calculations with the $I_h$ symmetry by Kunisada et al.[23] and Schmidt et al.[40] According to their calculation, about 80 singlet states exist below the first triplet excited state, and contribute to the specific heat below 10 K. This value is apparently in good agreement with our results, but the calculated specific heat strongly depends on the applied field as shown as solid and dotted curves in Fig. 3(a), which differs greatly from the experimental results. At 0 T, the clear peak at 2.4 K is apparent in the calculated specific heat. This peak increases and move toward lower temperatures with increasing magnetic field, which is clearly originated from the field dependent triplet excited states. According to their calculation, about 10 times triplet excited states of the singlet states exist at low energy and contribute to the specific heat below 10 K. The absence of the field dependent peaks in the experimentally obtained specific heat might be explained by the distribution of the triplet excited states. If the excited triplets are distributed in a wide energy range, the specific heat ascribed to the triplet excitations broadens and becomes insensitive to the applied field. As described in the previous subsection, the strong DM interactions can lead to the distribution of the triplet states even in the cluster with high symmetry.

### C. ESR

The difference of the distribution of singlet and triplet states between $\{Mo_{72}V_{30}\}$ and $\{W_{72}V_{30}\}$ was also confirmed by ESR measurements in the pulsed fields. The high-field ESR measurement is a powerful method to investigate the present polyhedral Kagomé clusters, because it can separate signals coming from the intrinsic clusters and the impurities by the difference of ESR linewidth and line shape. Additionally, it enables quantitative evaluation of the intrinsic intensity by comparison of the temperature dependence of signals between the intrinsic and the impurity spins. Although the singlet state is ESR-silent, it can be evaluated quantitatively by the change of the intensity in the triplet excitations. We used the method to analyze low-lying energy levels of polyhedron cluster explained in an earlier report.[21]

First, we present the temperature dependence of the ESR spectra for {Mo$_{72}$V$_{30}$} in Fig. 4. The employed frequency is 190 GHz. The inset of Fig. 4 shows that the two absorption curves are observed at 6.5 and 6.9 T. Hereinafter, we designate the broad and sharp resonances respectively as α and β. Signals α and β can be fitted by the sum of two Gaussian functions as indicated by the thin solid curve in the inset of Fig. 4. Each integrated intensity of α and β is obtained from fitted ESR spectra, where the temperature dependences of the integrated intensities for α (open squares) and β (solid circles) are presented in Fig. 5(a). The integrated intensity of β, which gradually increases concomitantly with decreasing the temperature, is well-fitted with the following function as

$$I_{\mathrm{imp}} = A N_{\mathrm{imp}} \tanh(g\mu_{\mathrm{B}} S \mu_0 H / k_{\mathrm{B}} T), \tag{1}$$

shown as a red dotted curve in Fig. 5(a) assuming the resonance from the $S = 1/2$ impurities. Here, $A$ and $N_{\mathrm{imp}}$ respectively denote the coefficient proportional to the power and frequency of the radiation and the number of impurities.[41] Therefore, β can be assigned to resonances deriving from the paramagnetic spins of the impurities. Thereby, α is an intrinsic resonance originating from {Mo$_{72}$V$_{30}$}. Presumably, the intrinsic signal α originates from the excited triplet states. The integrated intensity of the intrinsic signal ($I_{\mathrm{int}}$) is proportional to the population difference between the triplet states as

$$I_{\mathrm{int}} = A\{2(N_{\mathrm{int,M=-1}} - N_{\mathrm{int,M=0}}) + 2(N_{\mathrm{int,M=0}} - N_{\mathrm{int,M=+1}})\}. \tag{2}$$

The factor of 2 corresponds to the square of the transition matrix element, and $A$ is the same unknown coefficient used in Eq. (1). When intrinsic ESR intensity $I_{\mathrm{int}}$ is normalized with $AN_{\mathrm{imp}}$, the unknown coefficient

$A$ is eliminated, which engenders,

$$I'_{\text{int}} = \frac{I_{\text{int}}}{AN_{\text{imp}}} = \frac{N_{\text{int}}}{N_{\text{imp}}} \times \frac{2}{Z} \times \sum_{\text{triplet } i} \left\{ e^{-(E_i - g\mu_B\mu_0 H)/k_B T} - e^{-(E_i + g\mu_B\mu_0 H)/k_B T} \right\}, \quad (3)$$

where $Z$ denotes the partition function, and $AN_{\text{imp}} = 9.086 \times 10^{-5}$ is obtained by fitting of the temperature variation of β, and $N_{\text{int}}/N_{\text{imp}} \sim 1.14$ is estimated from the *M-H* curve. Equation (3) is useful because experimentally obtained $I'_{\text{int}}$ can be compared with the calculated $I'_{\text{int}}$ from the energy levels of the singlet and triplet states.

In Fig. 5(a), α drops at around 25 K with decreasing temperature. It becomes zero (i.e. signal is lost) at temperatures lower than 10 K. This result suggests that singlet states are realized below 10 K. The energy levels of {Mo$_{72}$V$_{30}$} are calculated using the following Heisenberg Hamiltonian with consideration of the Zeeman term as

$$H = \sum_{<i,j>} J_{ij} \vec{s_i} \cdot \vec{s_j} + g\mu_B\mu_0 H \sum_i s_i^z. \quad (4)$$

Here, $j_{ij}$ represents the exchange interaction between the spins at sites $i$ and $j$. Also, $\vec{s_i}$ and $s_i^z$ respectively represent the spin operator and z-component of a spin at site $i$. The eigenvalues of the Hamiltonian are calculated using the Lanczos method.[42,43] For simplicity, we consider only exchange interaction $J$ between the nearest-neighbor spins. Figure 5(b) shows the energy levels normalized by $J$ as a function of the total spin number *S*. Because of the high symmetry of the model, numerous singlet states exist near the ground state. Using these energy levels, the relative intrinsic intensity (i.e., $I'_{\text{int}}$) is calculated for the various $J$ values. Results for $J$ = 100, 200, 300, 500, 800 and 1000 K are shown as dotted and solid curves in Fig. 5(a). Figure 5(a) shows that the temperature dependences of the calculated $I'_{\text{int}}$ for all $J$ values are not reproducing the experimental results, which indicates that this single-$J$ model is unsuitable for {Mo$_{72}$V$_{30}$}. We should consider the model with the reduced symmetry, which is consistent with the results of the specific heat.

Next, we show the ESR spectra for {W$_{72}$V$_{30}$} in Fig. 6. The employed frequency is 405 GHz. The two absorption curves denoted by α and β are observed at around 14.8 T as shown in Fig. 6. These spectra can also be fitted by the sum of the two Gaussian functions at temperatures lower than 70 K, as indicated by the thin solid curves in Fig. 6. The integrated intensities of α and β are shown as functions of the temperature in Fig. 7. Both α and β gradually increases concomitantly with decreasing temperature. The integrated intensity of β is well-fitted by the Eq. (1) below 70 K as shown by dotted curve in Fig. 7. Therefore, it can be assigned to the resonances originating from the paramagnetic spins of the impurities. On the other hand, the integrated intensity of α deviates from the fitting curve below 10 K, which indicates that the temperature dependence of the integrated intensity of α cannot be explained solely by the paramagnetic impurity spins.

Here, we assume the existence of the intrinsic paramagnetic spins to explain the temperature dependence of α for $\{W_{72}V_{30}\}$ portrayed in Fig. 7. If the impurity spin couples with a spin in a $\{W_{72}V_{30}\}$ cluster, the singlet state partly breaks. Therefore, the extra-paramagnetic spins remain at low temperatures. Consequently, the increase of the integrated intensity of α can be attributed to the existence of the triplet spins in a $\{W_{72}V_{30}\}$ cluster coupling with the impurities at low temperatures. This contribution is also apparent in the ESR signal of β for $\{Mo_{72}V_{30}\}$ as asymmetric spectra, as portrayed in Fig. 4 (because the employed frequency of ESR measurement for $\{Mo_{72}V_{30}\}$ is lower than that of $\{W_{72}V_{30}\}$, the shoulder structure seen in Fig. 7 is not observed for $\{Mo_{72}V_{30}\}$).

It is noteworthy that no clear signal attributable to the triplet excited states is observed below 70 K for $\{W_{72}V_{30}\}$, as contrasted with $\{Mo_{72}V_{30}\}$. This fact suggests that numerous singlet states exist not only below the lowest triplet excited state but also above the lowest triplet excited state because of the high symmetry of $\{W_{72}V_{30}\}$ cluster [see Fig. 5(b)], which reduces the relative population of the triplet state (i.e., it leads to low ESR intensity of the triplet states). Therefore, the results of the ESR measurement support the specific heat data for both compounds.

## IV. Summary

We investigated the magnetic properties of $\{Mo_{72}V_{30}\}$ and $\{W_{72}V_{30}\}$ using magnetization, specific heat, and ESR measurements in pulsed and steady fields. Here, we summarize our conclusions as the following.

(1) The ground state in both {$Mo_{72}V_{30}$} and {$W_{72}V_{30}$} is a singlet.

(2) For {$Mo_{72}V_{30}$}, a few low-energy excited states, including triplet states, contribute to the field-dependent specific heat at low temperatures. The temperature and field variation of the specific heat demonstrate the validity of the theoretical model proposed by Kunisada *et al.*, which predicts that the icosidodecahedron of {$Mo_{72}V_{30}$} is distorted into the $D_{5h}$ symmetry. Therefore, it partly removes the low-lying singlet states.

(3) For {$W_{72}V_{30}$}, there are about 70 singlet excited states below 10 K, which contribute to the large and field-insensitive specific heat. Through comparison of the measured and calculated magnetic entropy, the high symmetry of the icosidodecahedron of {$W_{72}V_{30}$} is confirmed. No evidence exists of triplet excitations in the specific heat, although a large number of triplet excitations is predicted by the theoretical model with $I_h$ symmetry. This might indicate that the excited triplets are distributed in a wide energy range.

(4) The difference of the distribution of states between the two compounds with different symmetry is also confirmed by the ESR measurements in pulsed fields. For {$Mo_{72}V_{30}$}, the ESR signal originating from the triplet excited states is observed. The comparison of the experimentally obtained integrated intensity with the theoretical calculation confirmed the structural distortion of the icosidodecahedron of {$Mo_{72}V_{30}$}. On the other hand, no clear signal originating from the triplet excited states is observed below 70 K for {$W_{72}V_{30}$}, which suggests that numerous singlet states exist in a wide energy range because of the high symmetry of {$W_{72}V_{30}$} cluster.

(5) Monotonic *M-H* curves with no steps are observed in both {$Mo_{72}V_{30}$} and {$W_{72}V_{30}$}. We suggest the strong DM interaction can lead to the distribution of states, and therefore, the monotonic *M-H* curves. However, the discrepancy between the theory and the experiment is still remaining. Therefore, additional investigations must be conducted to elucidate the ground state and excited states in these materials.

## Acknowledgement

Some of this work was performed through joint research with the Institute for Solid State Physics of the University of Tokyo. This work was partly supported by the Ministry of Education, Culture, Sports, Science and Technology, Japan, through a Grant-in-Aid for Scientific Research (A) (Grant No. 20244052), a Grant-in-Aid for Early Career Scientist (Grant No. 18K13979) and a Grant-in-Aid for Scientific Research on Innovative Areas (Grant No. 24108704). This work was also partly supported by the International Collaboration Center, Institute for Materials Research, Tohoku University.

## References

[1] H.T. Diep, *Frustrated Spin Systems*, Second ed. (World Scientific, Singapore, 2013).

[2] O. Cépas and A. Ralko, Phys. Rev. B **84**, 020413(R) (2011).

[3] G.-W.W. Chern and R. Moessner, Phys. Rev. Lett. **110**, 077201 (2013).

[4] P. Lecheminant, B. Bernu, C. Lhuillier, L. Pierre, and P. Sindzingre, Phys. Rev. B **56**, 2521 (1997).

[5] C. Waldtmann, H.U. Everts, B. Bernu, C. Lhuillier, P. Sindzingre, P. Lecheminant, and L. Pierre, Eur. Phys. J. B **2**, 501 (1998).

[6] M. Mambrini and F. Mila, Eur. Phys. J. B **17**, 651 (2000).

[7] S. Yan, D.A. Huse, and S.R. White, Science **332**, 1173 (2011).

[8] H.J. Liao, Z.Y. Xie, J. Chen, Z.Y. Liu, H.D. Xie, R.Z. Huang, B. Normand, and T. Xiang, Phys. Rev. Lett. **118**, 137202 (2017).

[9] P. Mendels and F. Bert, Comptes Rendus Phys. **17**, 455 (2016).

[10] C. Lacroix, P. Mendels, and F. Mila, editors, *Introduction to Frustrated Magnetism* (Springer Berlin Heidelberg, Berlin, Heidelberg, 2011).

[11] A. Müller, A.M. Todea, J. Van Slageren, M. Dressel, H. Bögge, M. Schmidtmann, M. Luban, L. Engelhardt, and M. Rusu, Angew. Chemie - Int. Ed. **44**, 3857 (2005).

[12] B. Botar, P. Kögerler, and C.L. Hill, Chem. Commun. **21**, 3138 (2005).

[13] Y. Li, Y.G. Li, Z.M. Zhang, Q. Wu, and E.B. Wang, Inorg. Chem. Commun. **12**, 864 (2009).

[14] A.M. Todea, A. Merca, H. Bögge, T. Glaser, L. Engelhardt, R. Prozorov, M. Luban, and A. Müller, Chem. Commun. **0**, 3351 (2009).

[15] A. Müller, S. Sarkar, S.Q.N. Shah, H. Bögge, M. Schmidtmann, S. Sarkar, P. Kögerler, B. Hauptfleisch, A.X. Trautwein, and V. Schünemann, Angew. Chemie - Int. Ed. **38**, 3238 (1999).

[16] A.M. Todea, A. Merca, H. Bögge, J. Van Slageren, M. Dressel, L. Engelhardt, M. Luban, T. Glaser, M. Henry, and A. Müller, Angew. Chemie - Int. Ed. **46**, 6106 (2007).

[17] A. Müller, F. Peters, M.T. Pope, and D. Gatteschi, Chem. Rev. **98**, 239 (1998).

[18] C. Schröder, H. Nojiri, J. Schnack, P. Hage, M. Luban, and P. Kögerler, Phys. Rev. Lett. **94**, 017205 (2005).

[19] M.A. Palacios, E. Moreno Pineda, S. Sanz, R. Inglis, M.B. Pitak, S.J. Coles, M. Evangelisti, H. Nojiri, C. Heesing, E.K. Brechin, J. Schnack, and R.E.P. Winpenny, ChemPhysChem **17**, 55 (2016).

[20] J. Schnack and O. Wendland, Eur. Phys. J. B **78**, 535 (2010).

[21] Y. Oshima, H. Nojiri, J. Schnack, P. Kögerler, and M. Luban, Phys. Rev. B **85**, 024413 (2012).

[22] N. Kunisada, S. Takemura, and Y. Fukumoto, J. Phys. Conf. Ser. **145**, 012083 (2009).

[23] N. Kunisada and Y. Fukumoto, Prog. Theor. Exp. Phys. **2014**, 41I01 (2014).

[24] J. Schnack, A.M. Todea, A. Müller, H. Nojiri, S. Yeninas, Y. Furukawa, R. Prozorov, and M. Luban, ArXiv/Condmat 1304.2603v1 (2013).

[25] H. Nojiri, K.Y. Choi, and N. Kitamura, J. Magn. Magn. Mater. **310**, 1468 (2007).

[26] M. Motokawa, Reports Prog. Phys. **67**, 1995 (2004).

[27] S. Yamashita, Y. Nakazawa, M. Oguni, Y. Oshima, H. Nojiri, Y. Shimizu, K. Miyagawa, and K. Kanoda, Nat. Phys. **4**, 459 (2008).

[28] H. Nojiri, Y. Ajiro, T. Asano, and J.P. Boucher, New J. Phys. **8**, 218 (2006).

[29] A. Müller, M. Koop, H. Bögge, M. Schmidtmann, F. Peters, and P. Kögerler, Chem. Commun **3**, 1885 (1999).

[30] A. Müller, P. Kögerler, and A.W.M. Dress, Coord. Chem. Rev. **222**, 193 (2001).

[31] I. Rousochatzakis, Y. Ajiro, H. Mitamura, P. Kögerler, and M. Luban, Phys. Rev. Lett. **94**, 147204 (2005).

[32] J. Schnack, M. Luban, and R. Modler, Europhys. Lett. **56**, 863 (2001).

[33] E. Larry and L. Marshall, Dalt. Trans. **39**, 4687 (2010).

[34] C. Schröder, R. Prozorov, P. Kögerler, M.D. Vannette, X. Fang, M. Luban, A. Matsuo, K. Kindo, A. Müller, and A.M. Todea, Phys. Rev. B **77**, 224409 (2008).

[35] Y. Fukumoto, Y. Yokoyama, and H. Nakano, ArXiv/Condmat 1803. 06485v1 (2018).

[36] J. Schnack, Phys. Rev. B **93**, 054421 (2016).

[37] C. Heesing, Einfluss Der Dzyaloshinskii-Moriya-Wechselwirkung Auf Magnetische Moleküle, Ph. D. Thesis, Universität Bielefeld, 2016.

[38] A. Kramida, Y. Ralchenko, and J. Reader (NIST ASD Team), *NIST Atomic Spectra Database* (National Institute of Standards and Technology, Gaithersburg, MD, 2018). Available at https://www.nist.gov/pml/atomic-spectra-database.

[39] Z.D. Fu, P. Kögerler, U. Rücker, Y. Su, R. Mittal, and T. Brückel, New J. Phys. **12**, 083044 (2010).

[40] J. Schnack, J. Magn. Magn. Mater. **295**, 164 (2005).

[41] A. Abragam and B. Bleaney, *Electron Paramagnetic Resonance of Transition Ions* (Oxford University Press, 1970).

[42] C. Lanczos, J. Res. Natl. Bur. Stand. (1934). **45**, 255 (1950).

[43] J. Schnack, P. Hage, and H.-J. Schmidt, J. Comput. Phys. **227**, 4512 (2008).

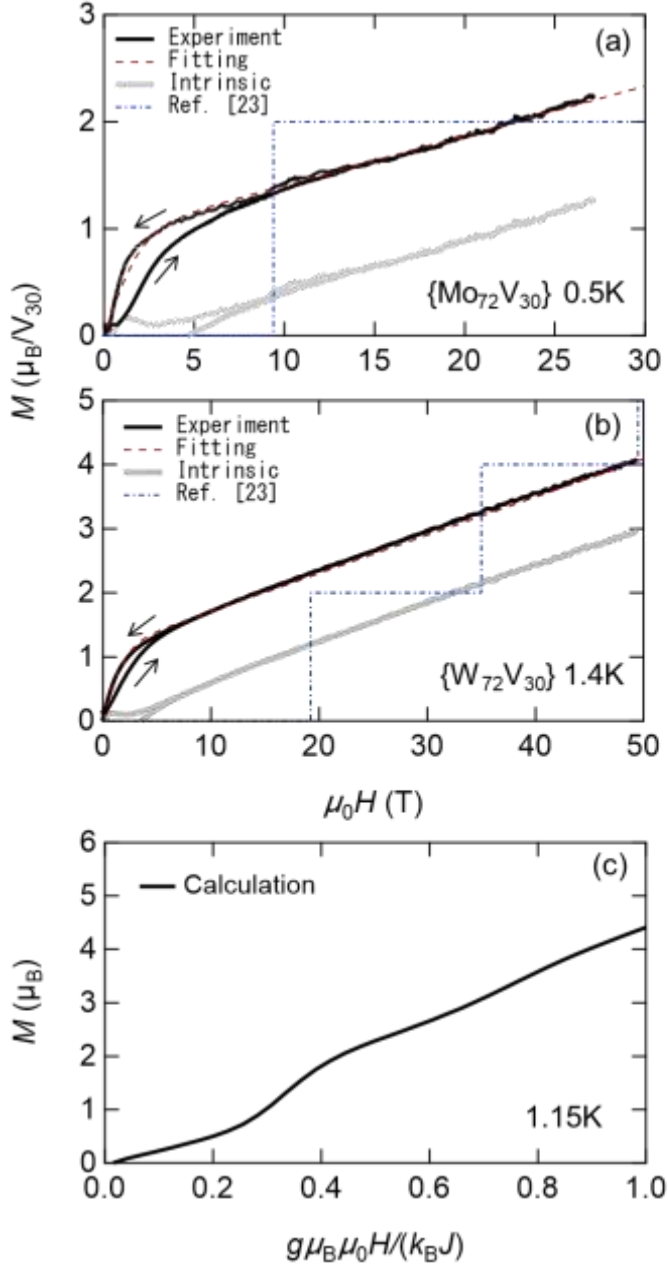

FIG. 1. (Color online) *M-H* curves measured in the pulsed magnetic fields (black solid curve): (a) {Mo$_{72}$V$_{30}$} at 0.5 K, (b) {W$_{72}$V$_{30}$} at 1.4 K. Red dashed curves show results of fitting by the sum of the linear term and Brillouin function for *S* = 1/2. Gray solid curves show the linear (intrinsic) parts of the magnetization. Blue dashed lines show results of the theoretical calculations performed by Kunisada *et al.*[23] (c) *M-H* curve calculated using the Heisenberg model considering DM interaction and the Zeeman effect. In the calculation, that corresponds to {W$_{72}$V$_{30}$} the DM vectors point radially outwards. The DM interaction strength is *J*/5 of the Heisenberg interaction *J*. The temperature was chosen as *T* = 1.15 K. Because of the complete lack of symmetry, very time consuming calculations were performed using the finite-temperature Lanczos method.[37]

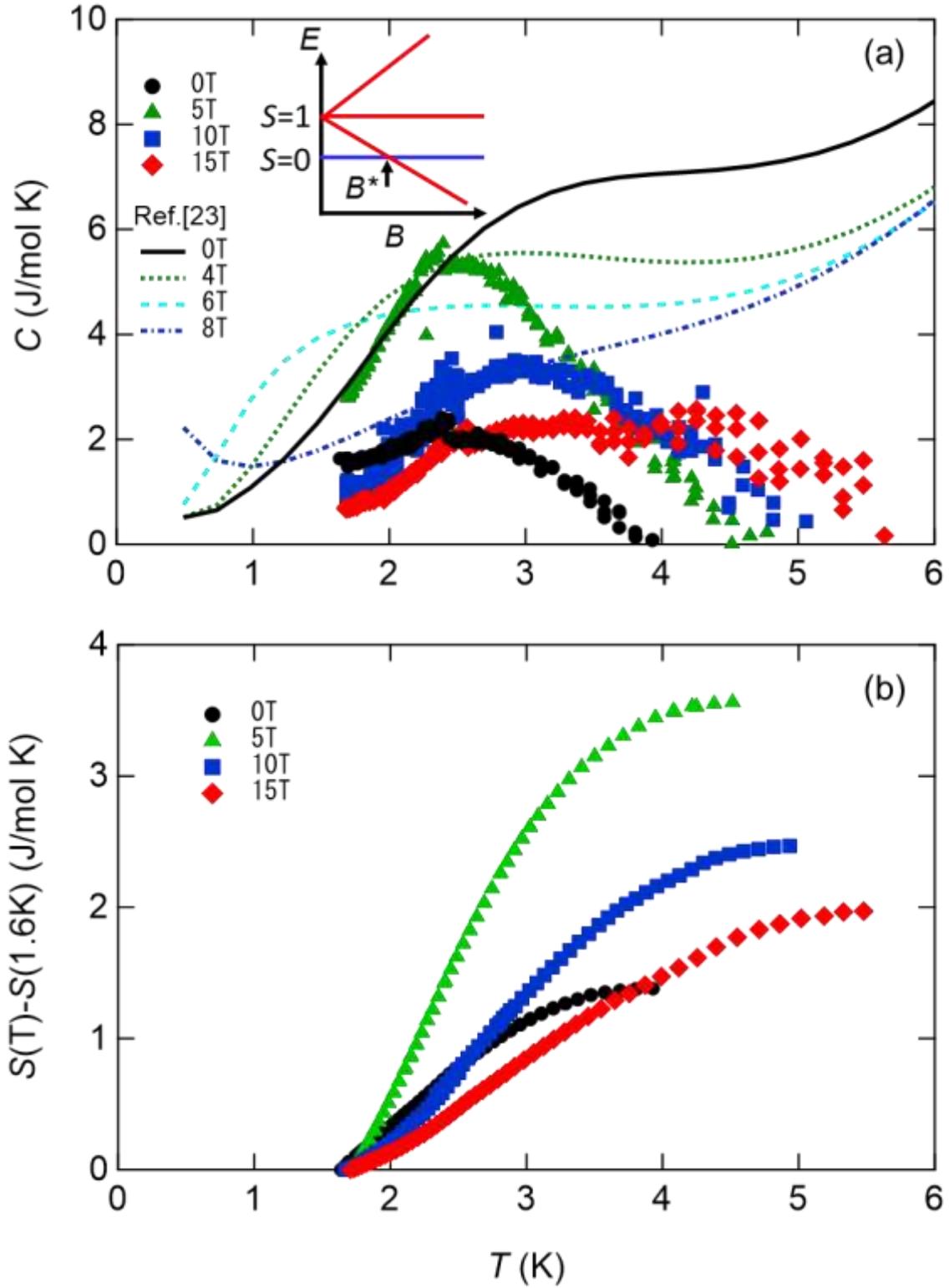

FIG. 2. (Color online) Temperature dependence of (a) specific heat and (b) entropy calculated from the specific heat at 0, 5, 10 and 15 T for the polycrystalline {$Mo_{72}V_{30}$}. Solid and dotted curves show the results of the theoretical calculations performed by Kunisada et al.[23]

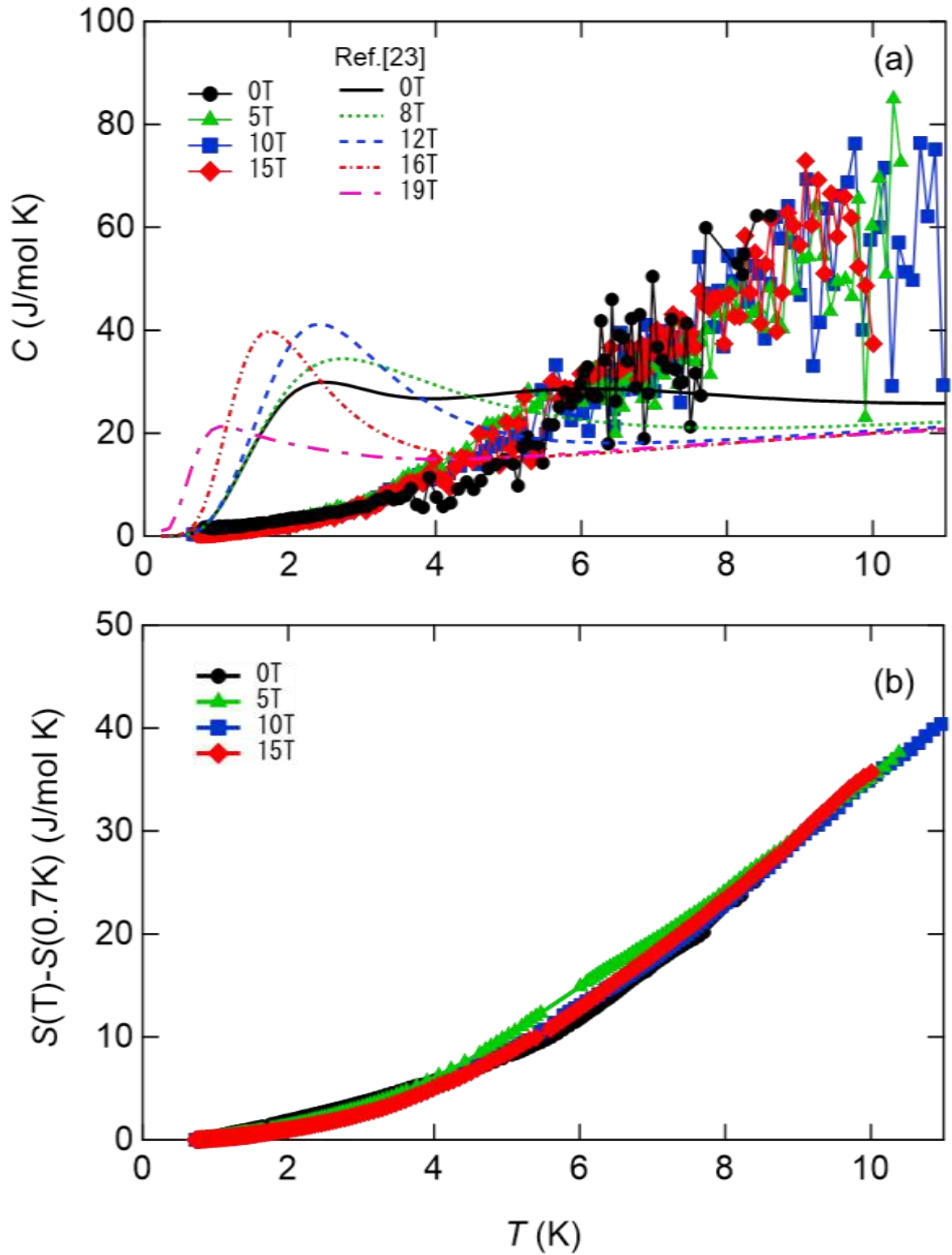

**FIG. 3.** (Color online) Temperature dependence of (a) specific heat and (b) entropy calculated from the specific heat at 0, 5, 10 and 15 T for the polycrystalline {$W_{72}V_{30}$}. Solid and dotted curves show results of theoretical calculations reported by Kunisada et al.[23]

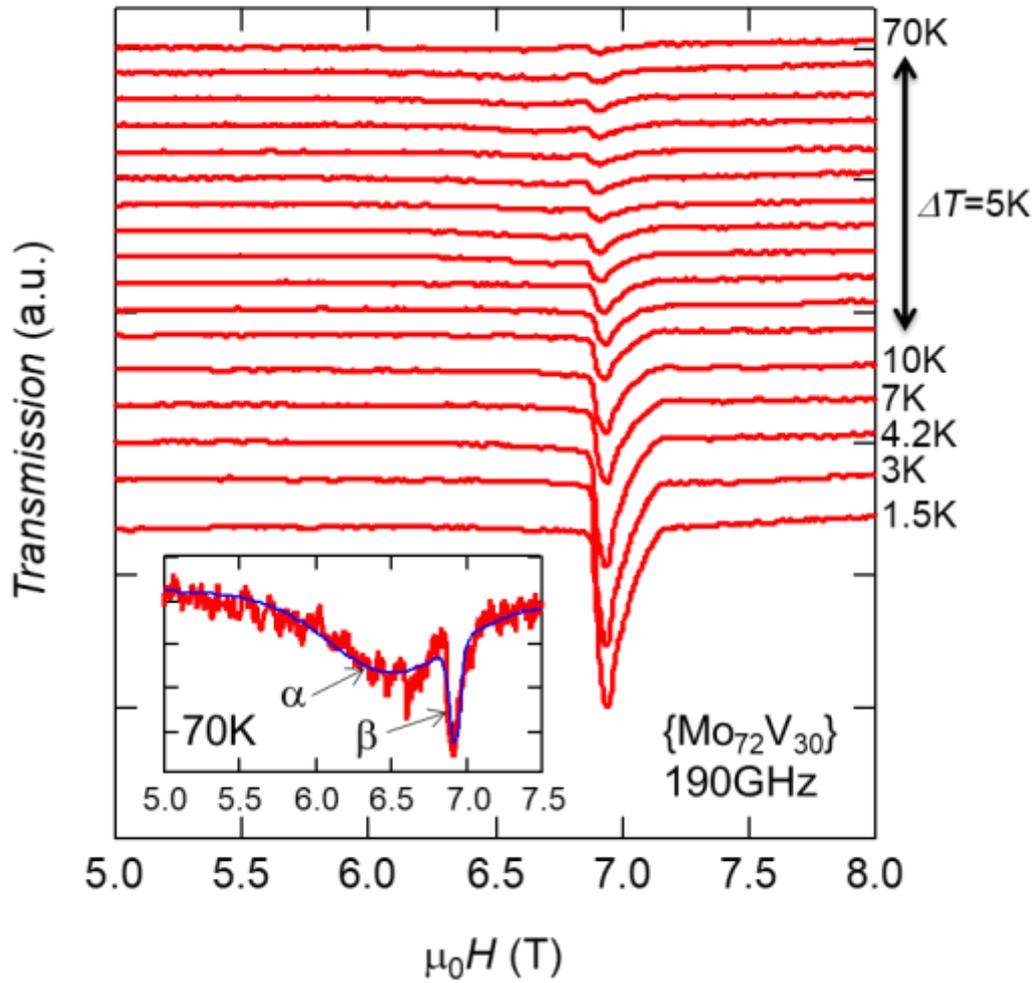

**FIG. 4.** (Color online) Temperature dependences of ESR spectra for 190 GHz in {Mo$_{72}$V$_{30}$}. The inset shows the spectrum of 70 K. The blue solid curve shows fitting by sum of the resonance α and β.

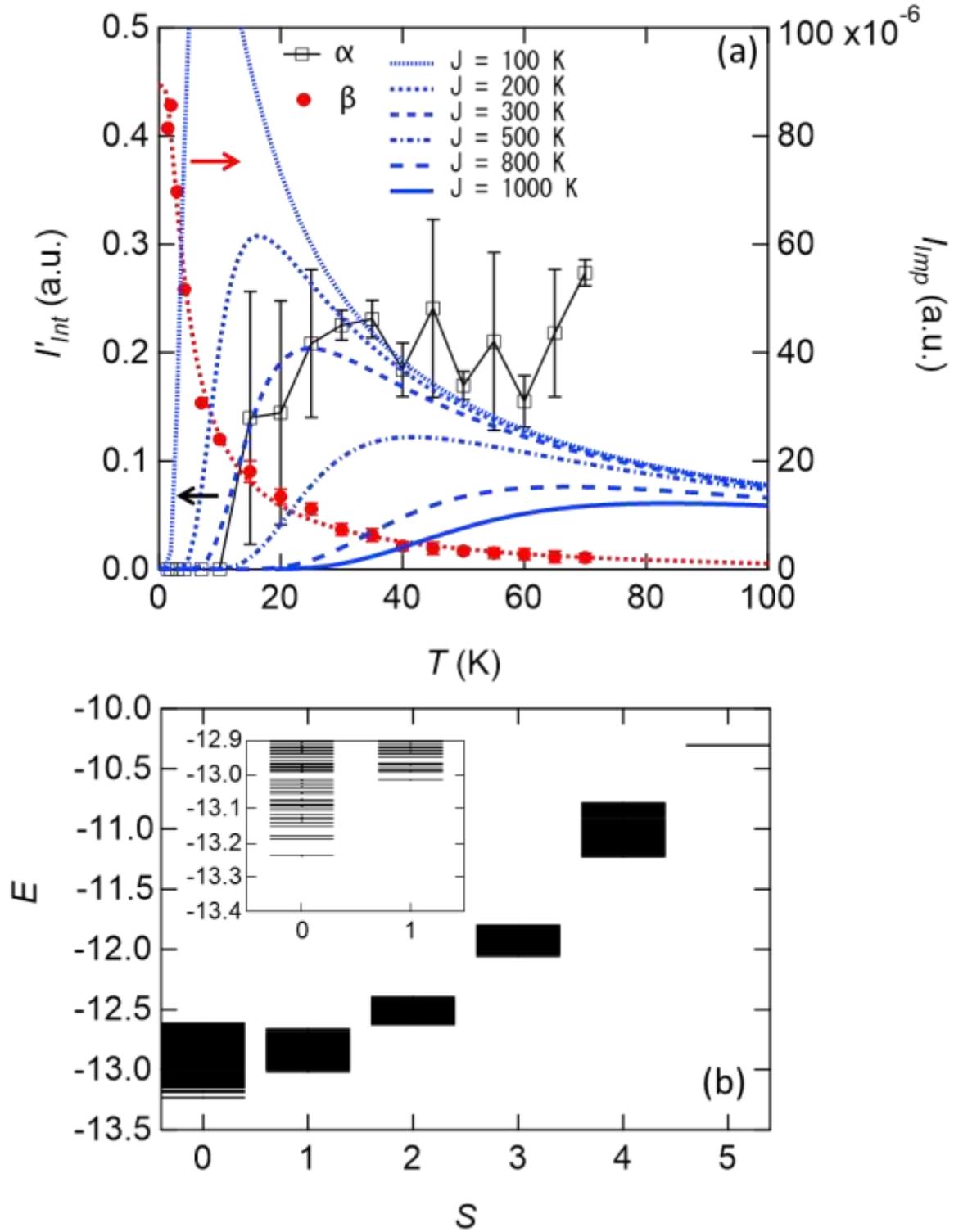

**FIG. 5.** (Color online) (a) Temperature dependence of the integrated intensities of ESR spectra for {Mo$_{72}$V$_{30}$}. α and β are explained in the inset of Fig. 4. α is normalized by $AN_{\text{imp}}$. (b) Low-lying energy levels normalized by $J$ as a function of total spin quantum number $S$. The inset is an enlarged view at around the ground state.

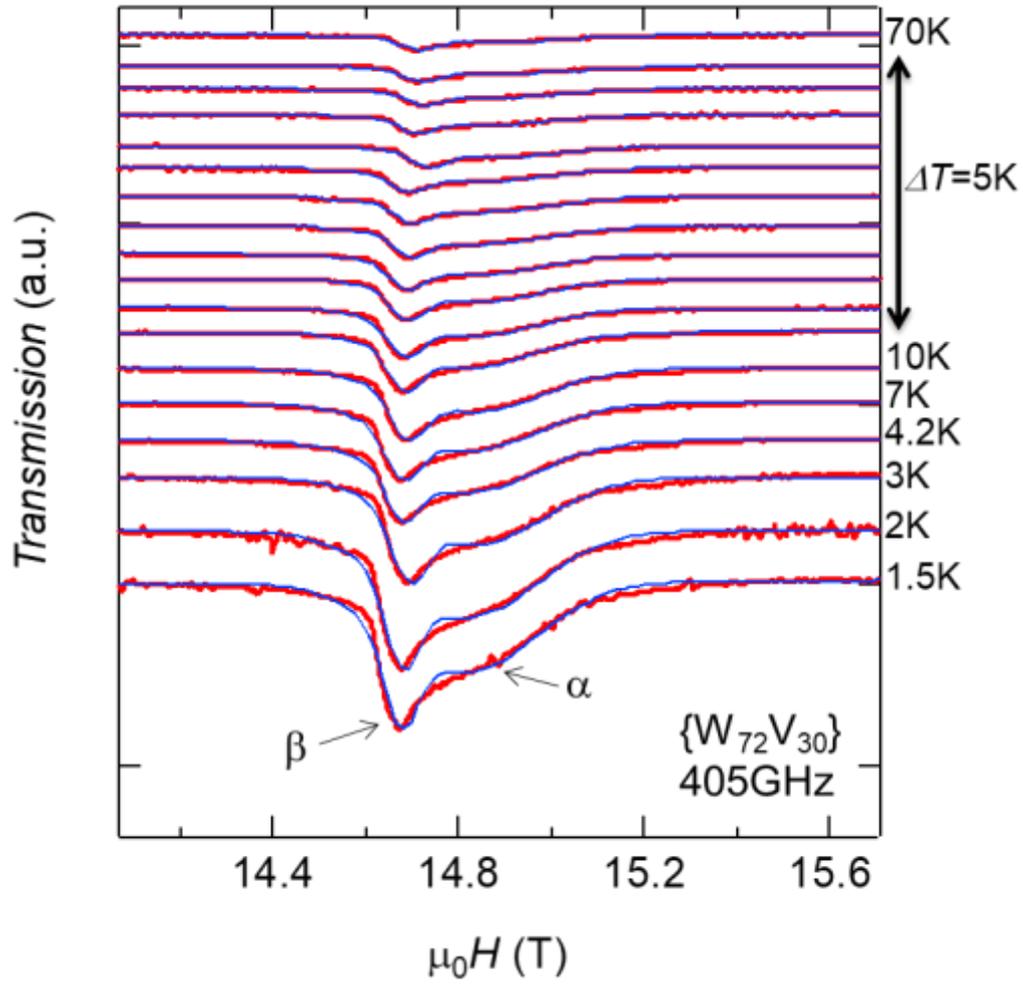

**FIG. 6.** (Color online) Temperature dependences of ESR spectra for 405 GHz in {$W_{72}V_{30}$}. The thin solid curve shows the fitting by the sum of the resonance α and β.

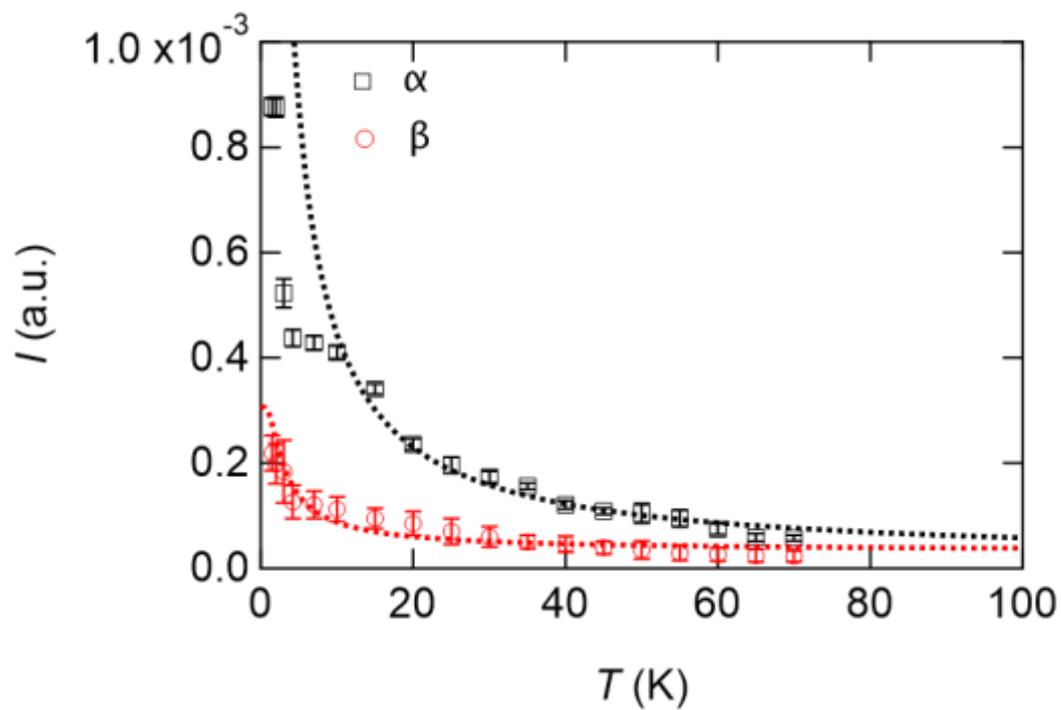

**FIG. 7.** (Color online) Temperature dependence of the integrated intensities for $\{W_{72}V_{30}\}$. α and β are explained in the inset of Fig. 6.